\documentclass[aps,pre,twocolumn,showpacs,epsfig,epsf,amssymb,floatfix]{revtex4-1}

\usepackage{graphicx}
\usepackage{dcolumn}
\usepackage{bm}

\begin{document}

\title{Emergent Softening and Stiffening Dictate Transport of Active Filaments}

\author{Bipul Biswas, Prasanna More and Hima Nagamanasa Kandula}
\affiliation{Physics Department, University of Massachusetts Amherst, Massachusetts, USA}

\maketitle
\textbf{Active semiflexible filaments are crucial in various biophysical processes, yet insights into their single-filament behavior have predominantly relied on theory and simulations, owing to the scarcity of controllable synthetic systems. Here, we present an experimental platform of active semiflexible filaments composed of dielectric colloidal particles, activated by an alternating electric field that induces contractile or extensile electrohydrodynamic (EHD) flows. Our experiments reveal that contractile flow generating filaments undergo softening, significantly expanding the range of accessible conformations, whereas filaments composed of extensile flow monomers exhibit active stiffening. By independently tuning filament elasticity and activity, we demonstrate that the competition between elastic restoring forces and emergent hydrodynamic interactions along the filament governs conformational dynamics. Crucially, we discover that the timescale of conformational dynamics directly governs transport behavior: enhanced fluctuations promote diffusion while stiffening facilitates directed propulsion of nonlinear filaments. Together, our direct visualization studies elucidate the links between inherent filament properties, microscopic activity, and emergent transport while establishing a versatile experimental platform of synthetic filamentous active matter.}

Filamentous structures in living matter are continuously subjected to out-of-equilibrium fluctuations driven by internal energy conversion processes, often referred to as active fluctuations. These active processes drastically influence the mechanics and dynamics of cytoskeletal filaments in cellular streaming \cite{suzuki2017spatial, ganguly2012cytoplasmi}, cilia and flagella in locomotion \cite{gilpin2020multiscale}, and chromatin organization driven by ATP \cite{zidovska2013micron, saintillan2018extensile}. In particular, the interplay between filament elasticity and activity gives rise to rich emergent behavior in semi-flexible active filaments, including novel mechanical responses \cite{de2019cytoskeletal, deblais2020rheology, heeremans2022chromatographic}, superfluid-like response to shear \cite{lopez2015turning}, and complex collective dynamics \cite{chakrabarti2022multiscale, vliegenthart2020filamentous, sanchez2012spontaneous}. Understanding how activity and inherent filament elasticity modulate the dynamics of semi-flexible filaments, specifically, their conformational changes and self-propulsion, is therefore essential to unraveling their emergent properties \cite{winkler2020physics, winkler2017active}. Nevertheless, systematic studies of individual active biological filaments remain highly challenging due to difficulties in isolating filaments, independently tuning activity and elasticity, and achieving the spatio-temporal resolution needed to track filament and monomer dynamics. As a result, their behavior has been primarily investigated through theoretical and computational approaches \cite{jayaraman2012autonomous, manna2019emergent, isele2015self, ghosh2014dynamics, chaki2023polymer, winkler2017active, winkler2020physics}. This gap underscores the need for synthetic active filament systems that enable independent control over activity and inherent filament properties. Such systems are critical not only for advancing fundamental understanding of active matter but also for guiding the design of functional, bio-inspired materials with tunable mechanical and transport properties. \cite{winkler2020physics}.\\\\
Colloidal filaments have emerged as a promising experimental platform for studying filamentous matter \cite{yang2018superparamagnetic, byrom2014directing, biswas2017linking, biswas2021rigidity, mcmullen2018freely, kumar2024emergent}. These filaments are chemically linked chains of micron-sized colloidal particles whose monomer properties and flexibility can be systematically tuned. Importantly, they allow direct visualization of filament conformations and dynamics at the monomer and filament length scale, simultaneously enabling the ability to provide unique insights into the physics of semiflexible systems. While recent efforts have demonstrated the self-propulsion of active filaments and shape transformations of colloidal filaments under external fields \cite{yang2020reconfigurable, haque2023propulsion, kuei2017strings, biswas2017linking, vutukuri2017rational, biswas2021colloidal, lyu2023biomimetic, kumar2024emergent}, an experimental system that enables direct and tunable control over both filament elasticity and internal active forces, crucial for investigating active semiflexible filaments, remains lacking \cite{jayaraman2012autonomous, laskar2013hydrodynamic, chelakkot2014flagellar, laskar2015brownian, manna2019emergent, mondal2020internal}.\\\\
To address this challenge, we develop active colloidal filaments composed of monomers that generate long-range hydrodynamic flows. Our monomer design is inspired by hydrodynamically active monomers that have been shown in theory and simulation studies to be key for replicating behaviors of biological filaments  \cite{saintillan2018extensile, jayaraman2012autonomous, laskar2013hydrodynamic, chelakkot2014flagellar, manna2019emergent, laskar2015brownian, mahajan2022self, winkler2020physics, winkler2017active}. In our filaments, each monomer autonomously generates electrohydrodynamic (EHD) flows in response to an applied alternating electric field (AC E-field), thereby making the colloidal filaments active. By tuning the properties of individual monomers, we control the nature of these flows to be either contractile or extensile \cite{ristenpart2007electrohydrodynamic, ma2015inducing, alvarez2021reconfigurable}. We show that the interplay between hydrodynamic coupling of monomers and filament elasticity governs the resulting dynamics. Interestingly, we find that the character of the EHD flows controls filament properties, with contractile flows inducing filament softening and enhancing conformational fluctuations, while extensile flows lead to active stiffening and shape locking. Crucially, the ability to independently control activity and rigidity reveals that filament transport is governed not merely by the shape of the instantaneous conformation, but also by the timescale of conformational fluctuations. Collectively, these findings provide new insights into the parameters that control the dynamics of active filaments and establish a versatile experimental platform with EHD flows as a powerful strategy for engineering synthetic active filamentous materials inspired by living active matter.
\begin{figure*}[ht]

\centering
\includegraphics[width=1\linewidth]{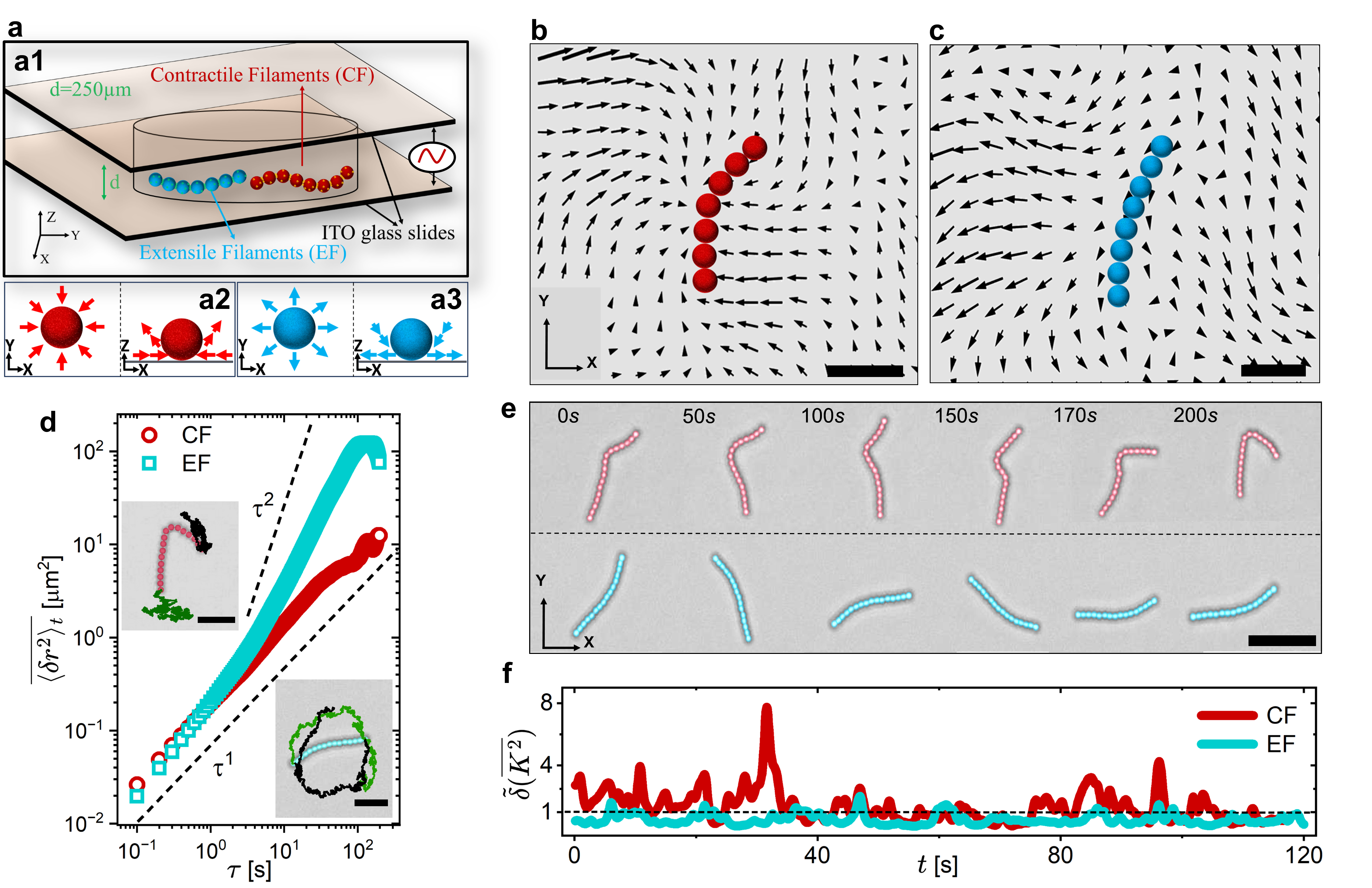}
\caption{\label{fig1}\textbf{Semi-flexible colloidal filaments made active with electrohydrodynamic (EHD) flows:} (a) - (a1) Schematic of the experimental setup: suspension of colloidal filaments sandwiched between two ITO electrodes to apply an orthogonal AC electric field; (a2) and (a3) shows schematic for single monomer generated contractile and extensile EHD flow profile, respectively. (b) and (c) show the EHD flow field generated by freely suspended contractile and extensile filaments, respectively, at $f = 0.7 kHz; E = 0.024V/\mu m $. The mean flow field is computed using $\mu$PIV for a total of $t=3.13s$ (100 frames) for the contractile filament (b), and $t=6.25s$ (200 frames) for the extensile filament (c). For (b) and (c), the scale bar is $3\mu m$. (d) Monomer averaged mean square displacement,  $\overline{ \langle \delta r^2\rangle_{t}}$ for a contractile filament (CF) (red circles) and an extensile filament (EF) (blue squares) at $f = 1 kHz; E = 0.024V/\mu\text{m} $. The lines of slope 1 and 2 are drawn as a guide to the eye. The inset shows the trajectories of end monomers of the CF (red) and the EF (blue). Scale bar $=5 \mu m$. (e) Optical microscopy snapshots of filament conformations of the CF (top panel) and EF (bottom panel) for filaments shown in Fig. \ref{fig1}d at $f = 1 kHz; E = 0.024V/\mu\text{m} $. CF exhibits conformational changes, while EF displays a fixed conformation and shows rigid body-like rotations.  Scale bar $=5\mu m$. (f) Fluctuations in the mean of local curvature squared, $\tilde{\delta}(\overline{K^2}) = \delta (\overline{K^2}) / \delta^{0} (\overline{K^2}) $, where $\delta (\overline{K^2}) = |\overline{K^2}(t)-\langle \overline{K^2} (t) \rangle_{t}|$ is normalized with $\delta^{0} (\overline{K^2})$ corresponding to the Brownian counterpart. CF (red) and EF (blue) demonstrate enhanced and suppressed conformational changes compared to their Brownian counterparts, respectively. Here, $\delta (\overline{K^2})$ and $\delta^{0} (\overline{K^2})$ values are smoothened using the Gaussian method over 20 frames (2$s$) interval. A dotted line at $\tilde{\delta}(\overline{K^2})=1$ ($\delta (\overline{K^2}) = \delta^{0} (\overline{K^2}$)) is drawn as a guide to the eye.
} 
\end{figure*}
\section{Semiflexible active colloidal filaments design}
\begin{figure*}[ht]

\centering
\includegraphics[width=1\linewidth]{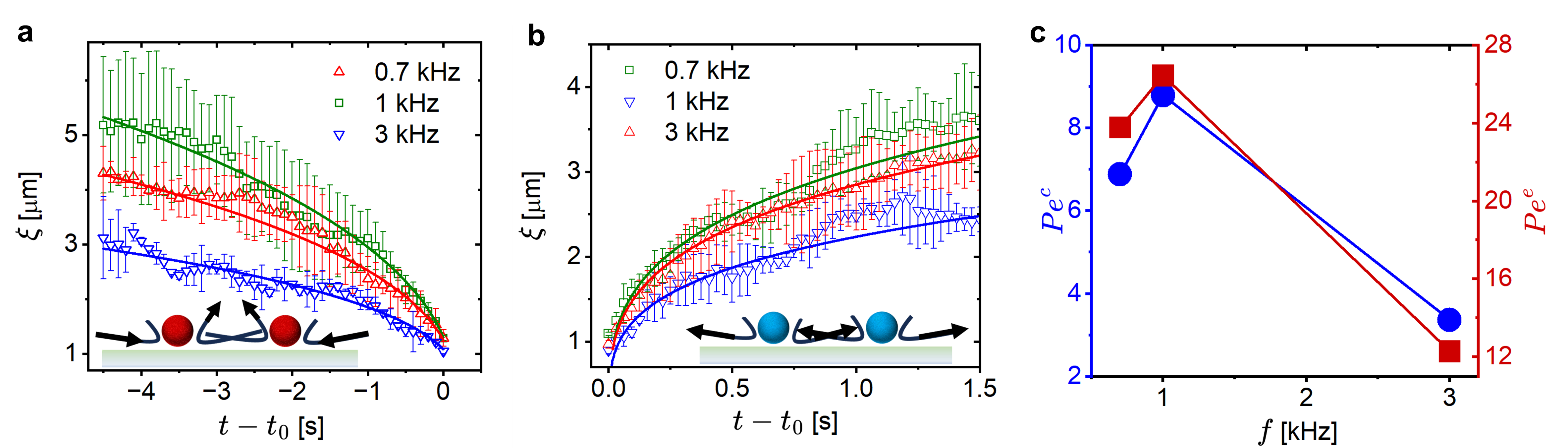}
\caption{\label{fig2} \textbf{Estimating $P\acute{e}clet$ number ($Pe$) from experiments:} $Pe$ is extracted from the effective attractive and repulsive interactions of contractile and extensile monomers, respectively. (a) and (b) show the separation between pairs of monomers $\xi$ as a function of time ($t-t_0$, where $t_0$ is the time at which the two monomers are in contact with each other) for contractile and extensile monomers, respectively. Error bars are standard deviations obtained from analyzing multiple pairs of monomers for each $f$. Solid lines are the power fits: $\xi \propto (t-t_0)^{n}$, with $n_c$ and $n_e$ being the fit exponents for contractile and extensile monomers. The extracted fit exponents for different $f$'s are: for 0.7 $kHz$ - $n_c \approx 0.42$ and $n_e \approx 0.28$; 1 $kHz$ - $n_c \approx 0.44$ and $n_e \approx 0.27$; 3 $kHz$ - $n_c\approx 0.34$ and $n_e\approx 0.25$. 
(c) The $Pe$ number as a function of $f$ at $E=0.024 V/ \mu m$ for contractile EHD denoted by $Pe^c=(|u_c(\xi=1)|*l_c)/D$ (blue; left axis), and extensile EHD represented by $Pe^e=(|u_e(\xi=1)|*l_c)/D$. $Pe$ is computed using the velocities obtained from the first derivative of fits in (a) and (b) at $\xi=1 \mu m$, average monomer separation in the colloidal filaments (Supplementary Figure S3). For easier comparison, we normalize them by $D/l_c$. Here, $D$ is the diffusion coefficient of a Brownian monomer, and $l_c$ is the characteristic length scale, which is considered as ($l_c \approx 1 \mu m$), the average separation between monomers in a filament.}

\end{figure*}

We fabricate chemically cross-linked semi-flexible colloidal filaments with a persistence length-to-contour length ratio ($L_p/L_C$) in the range of $1 -3$\cite{yang2018superparamagnetic,yang2020reconfigurable,biswas2017linking} (Methods, Supplementary Figure S1, Supplementary Movies 1-3). Colloidal filaments are made of polystyrene particles which due to their dielectric nature generate electrohydrodynamic (EHD) flows when subjected to an orthogonal alternating electric field (AC E-field) (Fig.\ref{fig1}a1). We can change the direction of the EHD flow by modulating the monomer zeta potential, $\zeta_p$, via the surface modification chemistry and/or solvent pH \cite{ma2015inducing, yang2019impact, alvarez2021reconfigurable} (Supplementary Note 3, Supplementary Table 1). Specifically, we develop two distinct surface functionalizations and corresponding cross-linking protocols to create filaments whose monomers produce either sink-like (\textbf{\textit{contractile}}) (Fig.\ref{fig1}a2, Supplementary Figures S2a and S2c) or source-like (\textit{\textbf{extensile}}) flows (Fig.\ref{fig1}a3, Supplementary Figures S2b and S2d) in the X-Y plane. In 3D, the flows are circulatory and hence the flows conserve mass despite being monopolar in 2D \cite{ristenpart2007electrohydrodynamic,yang2019impact}. However, as colloidal filaments settle to the bottom of the observation chamber, in-plane monopolar flows and their interactions dominate the dynamics. Accordingly, we classify the filaments as contractile filaments (CFs) and extensile filaments (EFs) based on their EHD flows. Micro-particle image velocimetry ($\mu $PIV) confirms the EHD flow profiles generated by the two filament types. In DI water (pH $\approx 6$), filaments made of $1,4$-dithiothreitol (DTT) functionalized monomers ($\zeta_p \approx -25mV $) generate contractile flows with arrows pointing inward toward the filament (Fig.\ref{fig1}b, Supplementary Movie 4) and those with polyethyleneimine (PEI) functionalized monomers ($\zeta_p \approx +40 mV$) produce extensile flows with arrows pointing outward away from the filament (Fig.\ref{fig1}c, Supplementary Movie 4).\\\\ 
We quantify filament dynamics using monomer averaged mean squared displacement (MSD), $\overline{\langle\delta r_{i}^2\rangle_{t}}  =(1/N) \Sigma_{i=1}^{N}(\langle (x_{i}(t)-x_{i}(t+\tau))^2\rangle_{t} + \langle (y_{i}(t)-y_{i}(t+\tau))^2\rangle_{t})$, where $\tau$ is the lagtime, $x_{i}(t)$ and $y_{i}(t)$ are the instantaneous centroid of the $i^{th}$ monomer, and $N$ is the total number of monomers in the filament. As indicated by the slope of the MSD on a log-log scale, the CF exhibits diffusive behavior  (slope $\approx 1$) while EF displays super-diffusion (slope $> 1$) at long times (Fig.\ref{fig1}d). This difference in their dynamics is also evident in the end-monomer trajectories: CFs exhibit random motion, while EFs follow circular trajectories, consistent with persistent rotation (insets in Fig.\ref{fig1}d). Strikingly, snapshots of the conformations reveal contrasting behaviour between the CF and EF (Fig.\ref{fig1}e). CFs exhibit frequent shape changes over the entire duration of the experiments, whereas EFs retain a stable conformation and behave as rigid rotating bodies. To quantify the conformational changes of the active filaments compared to their thermal counterparts, we examine fluctuations in the mean local curvature squared, $\tilde{\delta} (\overline{K^2}) = \delta (\overline{K^2}) / \delta^{0} (\overline{K^2}) $. Here $\delta (\overline{K^2}) = |\overline{K^2}(t)-\langle \overline{K^2} (t) \rangle_{t}|$, where $\overline{K^2} =\langle(d\theta / ds)^2\rangle$ averaged over the contour length with $d\theta / ds$ being the curvature measured over $ds = 2 \mu m$ which corresponds to three monomers. $\delta^{0} (\overline{K^2})$ is the average value of curvature in the curvature of the Brownian counterparts of the active filaments. For CF, we find that $\tilde{\delta} (\overline{K^2}) >1 $ and shows large variations indicative of greater conformational changes beyond those possible with thermal fluctuations (Fig.\ref{fig1}f - red). In contrast, EF shows minimal variation in fluctuations with  $\tilde{\delta} (\overline{K^2}) <1$, suggesting a suppression of conformational changes (Fig.\ref{fig1}f - blue). These surprisingly distinct observations with the two kids of EHD flows demand the determination of the direct correlation between the nature of the hydrodynamic flow and the emergent filament dynamics.\\\\ 
To this end, first, we determine the interaction between the monomers induced by the EHD flows. We find that contractile monomers exhibit attractive interactions, while extensile monomers repel each other when in close proximity (Supplementary Movie 5). We quantify the resulting active forces between monomers by defining $P\acute{e}clet$ number ($Pe$), $ Pe=(|u_{eff}(\xi)|*l_c)/D$. Here, $\xi$ is the separation between pairs of monomers (Fig.\ref{fig2}a and b), $D$ is the diffusion coefficient of a Brownian monomer, $l_c \approx 1 \mu m$ which is the average separation between monomers in the filament, and $u_{eff}(\xi)$ the effective velocity with which pairs of contractile monomers attract ($u_{eff}(\xi)=u_{c}(\xi)$) and extensile monomers repel ($u_{eff}(\xi)=u_{e}(\xi)$) (Supplementary Figure S3). We can control $Pe$ and thus the activity, by tuning the frequency of the applied AC electric field. Figure \ref{fig2}c shows $P{e}$ versus $f$ for $\xi = 1 \mu m$, the average bond length in our colloidal filaments. Surprisingly, $Pe$ exhibits a non-monotonic dependence on field frequency $f$ (Fig. \ref{fig2}c), in contrast with the monotonic increase expected for single-monomer EHD flow strength \cite{ristenpart2007electrohydrodynamic,yang2019impact,alvarez2021reconfigurable}(Supplementary Figure S2e). This difference likely arises from the competing effects of EHD flow strength and AC E-field-induced viscous hindrance (Supplementary Note 4, Supplementary Fig. S4).
\section{Activity induced conformational changes}
\begin{figure*}[ht]

\centering
\includegraphics[width=1\linewidth]{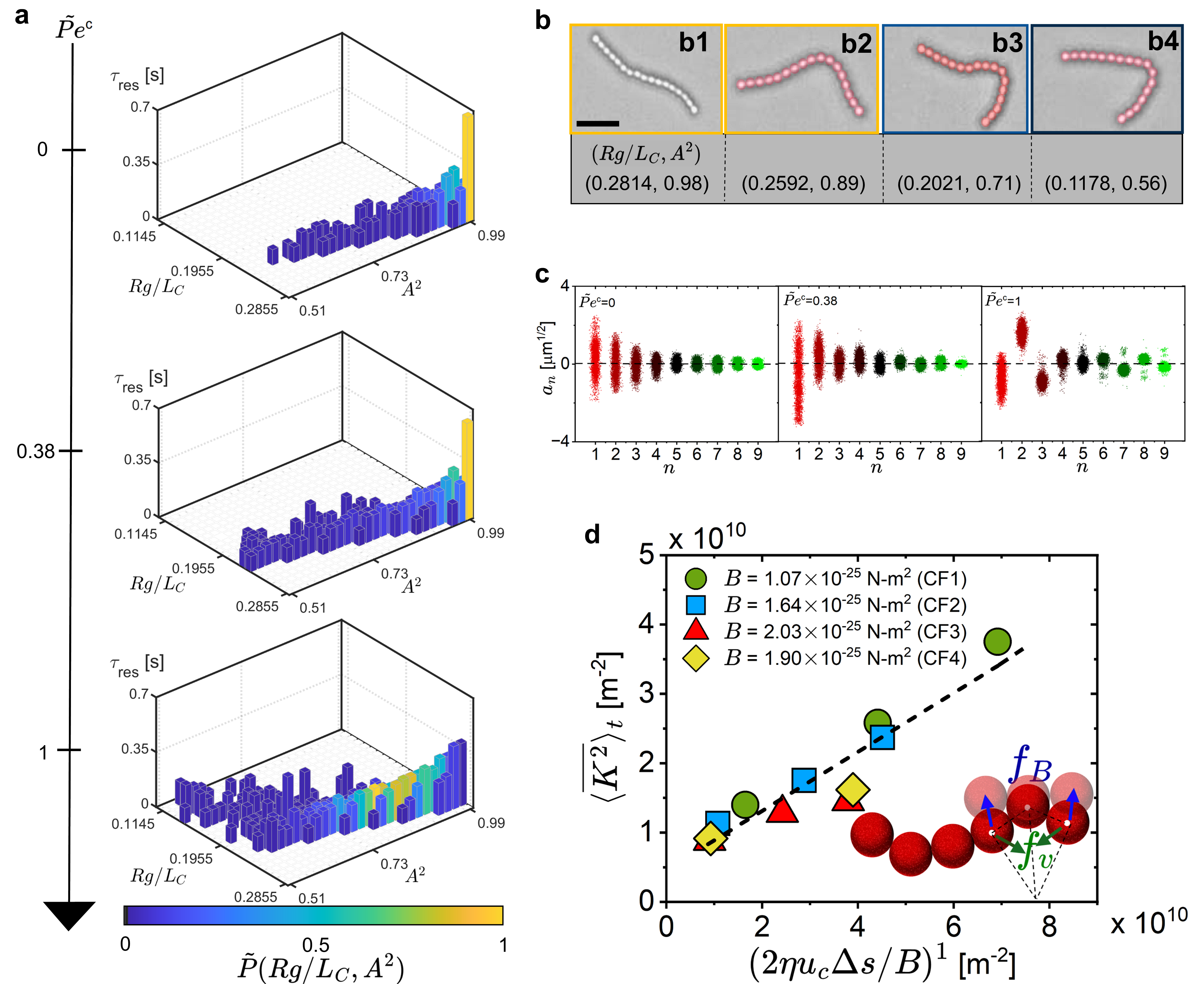}
\caption{\label{fig3} \textbf{Activity induced conformational changes and bending deformations of contractile filaments (CFs):} (a) Conformational maps of all the conformations sampled by CF1 at various normalized contractile $Pe$ number, $\tilde{Pe^c}$. The conformational space, $\mathbb{C}_{n_{1} \times n_{2}}$ defined by $Rg/L_C \in [0.11,0.29]$ (x-axis) and $A^2 \in [0.50,1.00]$ (y-axis), is populated by dividing it into $ (n_{1} \times n_{2})$ bins with x-bin width $=0.009$, and y-bin width $=0.02$ at all $\tilde{Pe^c}$. The color bar is the normalized probability, $\tilde{P(Rg/L_C, A^2)}=P(Rg/L_C, A^2)/P_{max}$, of finding the filament in a given conformation box $(Rg/L_C, A^2)$. The z-axis represents the average residence time, $\tau_{res}$, which is the average time spent by the filament in a box before hopping to another box of the $\mathbb{C}$. (b) Microscopy snapshots of representative conformations of CF1. $Rg/L_C$ and $A^2$ values for each conformation are written below (scale bar=$5 \mu m$). b1 is the conformation corresponding to the most probable conformation of CF1 at $\tilde {Pe^c}=0$, and b2-b4 correspond to the CF1 conformations at $\tilde {Pe^c}=1$. In all b1-b4, the color of the boundary is matched with its $\tilde{P(Rg/L_C, A^2)}$ color bar on $\mathbb{C}$. (c) The bending mode amplitudes, $a_n$ for CF1 for various $n$ at different $\tilde {Pe^c}$. The distribution of $a_n$ shows that modes of all length scales are affected by EHD flows. (d) Scaling analysis for CFs with different bending moduli ($B$). The mean of local curvature squared, $\overline{K^2}$ averaged for all conformations spanned by the CF, $\langle \overline{K^2} \rangle_{t}$, plotted against $(2\eta u_c \Delta s / B )^1$, where $\eta u_c \Delta s$ is contractile force for a length scale of $\Delta s$. As a minimum of three monomers are required for a local bend, we choose $\Delta s = 2 \mu m$. The data collapses onto a straight line for CFs with different $B$. The dotted line is a linear fit showing the agreement with equation \ref{e1}. The inset shows a schematic for the forces involved in a CF, contractile active force, $f_v$ (green), responsible for bending the filament locally competing against elastic restoring force, $f_B$ (blue), which tends to restore the filament to a locally straight conformation. The faint monomers in the figure represent the shape that purely elasticity wants to impose on the filament.} 

\end{figure*}

We next investigate the conformational dynamics of the two filament types, beginning with contractile filaments, which exhibit greater shape fluctuations than their extensile counterparts (Fig.\ref{fig1}d).
\textit{\textbf{Filaments with contractile monomers, CFs}}: 
To analyze conformational dynamics, we first identify the distinct shapes adopted by the filaments and the transitions between them. Each instantaneous conformation is characterized using two key parameters: the normalized radius of gyration $R_g/L_C$ and the acylindricity $A^2$. $Rg$ is the radius of gyration, which quantifies the spatial extent of the filament configuration, $L_C$ is the contour length, and $A^2$ measures the deviation from circular profile. These parameters are derived from the two-dimensional gyration tensor\cite{anderson2022polymer,liu2018morphological}, $G_{x y} = 1/N\sum_{i=1}^{N} (x_i-x_{cm}) (y_i-y_{cm})$, where $N$ is the number of monomers, $(x_i,y_i)$ coordinates of the centroid of the $i^{th}$ monomer and $(x_{cm},y_{cm})$ the center of mass of the filament in any given conformation. Using the real eigenvalues, $(\lambda_1,  \lambda_2)$ of $G_{x y}$, we compute $R_g^2 = \lambda_1 + \lambda_2$, and $A^2 = (\lambda_2 - \lambda_1)/R_g^2$. Subsequently, we construct a $2-D$ discretized conformational space, $\mathbb{C}_{n_{1} \times n_{2}}$. In this space, typically for semi-flexible filaments higher $R_g/L_C$ and $A^2$ correspond to linear conformations, while smaller values reflex compact conformations. Figure \ref{fig3}a illustrates $\mathbb{C}_{n_{1} \times n_{2}}$ of CF1, for varying normalized activity, $\tilde {Pe^c}=Pe^{c}/Pe^{c}_{max}$ with the color map indicating the probability $\tilde{P}(Rg/L_C, A^2)$. In the absence of activity ($\tilde {Pe^c}$ =0), CF1 primarily samples linear conformations (top panel \ref{fig3}a and Fig.\ref{fig3}b1). As $\tilde {Pe^c}$ increases, the area sampled on $\mathbb{C}$ increases indicating that the filament explores a broader range of conformations, with a shift in the most probable state toward more compact configurations represented by the smaller values of $R_g/L_C$,$A^2$ (snapshots in Fig.\ref{fig3}b2-b4 at $\tilde{Pe^c}=1$, Supplementary Movie 6), indicating that contractile activity drives the filament to compact states.\\\\ 
Subsequently, we quantify the dynamics of conformational transitions, we calculate the average residence time, $\tau_{res}$ defined as the average duration a filament remains within a given bin of the conformational space $\mathbb{C}$ before transitioning to another. In the absence of activity ($\tilde {Pe^c}=0$), very few compact conformations are explored and the small number of compact conformations explored also have significantly lower $\tau_{res}$ compared to linear conformations. This reflects the tendency of compact conformation to restoration to the linear conformation due to the elastic restoring forces. However, for CF1 under active forces $\tilde {Pe^c}>0$, $\tau_{res}$ for both the linear ($\tau_{res}\approx 0.35 s$) and most-probable compact conformations ($\tau_{res}\approx 0.22 s$) are comparable indicating that activity increases the stability of the compact. Furthermore the comparable values of $\tau_{res}$ for compact and linear conformations indicates that, even in the presence of activity, elasticity remains a key regulator of filament dynamics. Without elastic restoring forces, we would expect compact conformations to dominate with significantly longer $\tau_{res}$ and linear states to be significantly short-lived with negligible $\tau_{res}$. Therefore, the observed residence times reflect a balance between contractile active forces promoting compaction and inherent filament elasticity favoring extension (Fig.\ref{fig3}a). \\\\ 
Although the range of the contractile force generated by each monomer is relatively modest $\sim 4-5 \mu m$ (Fig. \ ref {fig2}), larger than monomer size and smaller compared to $L_p$, it influences the emergent conformations and fluctuations of filaments across multiple length scales. To quantitatively assess this, we decompose filament conformations into Fourier cosine modes by expanding the tangent angles $\theta(s)$  along the arc length $s$ \cite{fakhri2009diameter,li2010bending,stuij2021revealing} as $\theta(s)=\sum_{n=0}^{\infty} \theta_{n}(s)=\sqrt{\frac{2}{L_C}}\sum_{n=0}^{\infty}a_{n}cos(\frac{n\pi s}{L_C})$, where $n$ is the mode number and $a_n$ is the amplitude of the $n^{th}$ mode. In the absence of activity ($\tilde {Pe^c} = 0$), the distribution of $a_n$ is symmetric and centered around zero for all $n$, indicating that the mean conformation is a linear one (Fig.\ref{fig3}c). As activity increases, however, CF1 exhibits increasingly asymmetric distribution of $a_n$ for all $n$ signaling significant deviations from a linear shape. Low mode numbers ($n<4$) capture long wavelength deformations suggesting the mean conformation to be a compact one. Whereas deviation at high mode numbers ($n > 4$) reflect the enhanced local bending deformations (Supplementary Note 6, Supplementary Figure S6). This multiscale deformation is necessary for broadening of conformational space explored by active CFs (Fig.\ref{fig3}a). Furthermore, the observed spread in $a_n$ even for maximum activity possible $\tilde {Pe^c}=1$, indicates that active CFs continue to undergo shape fluctuations, maintaining a stochastic and a fluctuating steady state  (Supplementary Note 7, Supplementary Figure S7, Supplementary Movies 1 and 6). Unlike smooth, deterministic S- or C-shaped conformations commonly reported in systems dominated by active forces with negligible thermal and/or elastic contributions \cite{kumar2024emergent,yang2020reconfigurable,kuei2017strings} our CFs exhibit rugged conformations with fluctuations across scales. This behavior of our filaments results from the interplay of thermal fluctuations, elasticity, and EHD-driven activity—paralleling stochastic steady states seen in living systems such as MDCK-II primary cilia \cite{battle2016broken}. To further illustrate the role of elasticity in preserving this stochasticity, we present a control experiment using filaments that have a few bonds with weakened cross-links. These filaments undergo collapse into a maximally compact with no fluctuations in shape, a deterministic conformation, confirming that the absence of sufficient elastic restoring forces results in a fixed steady state (Supplementary Movie 7).\\\\
\begin{figure*}[ht]
\centering
\includegraphics[width=1\linewidth]{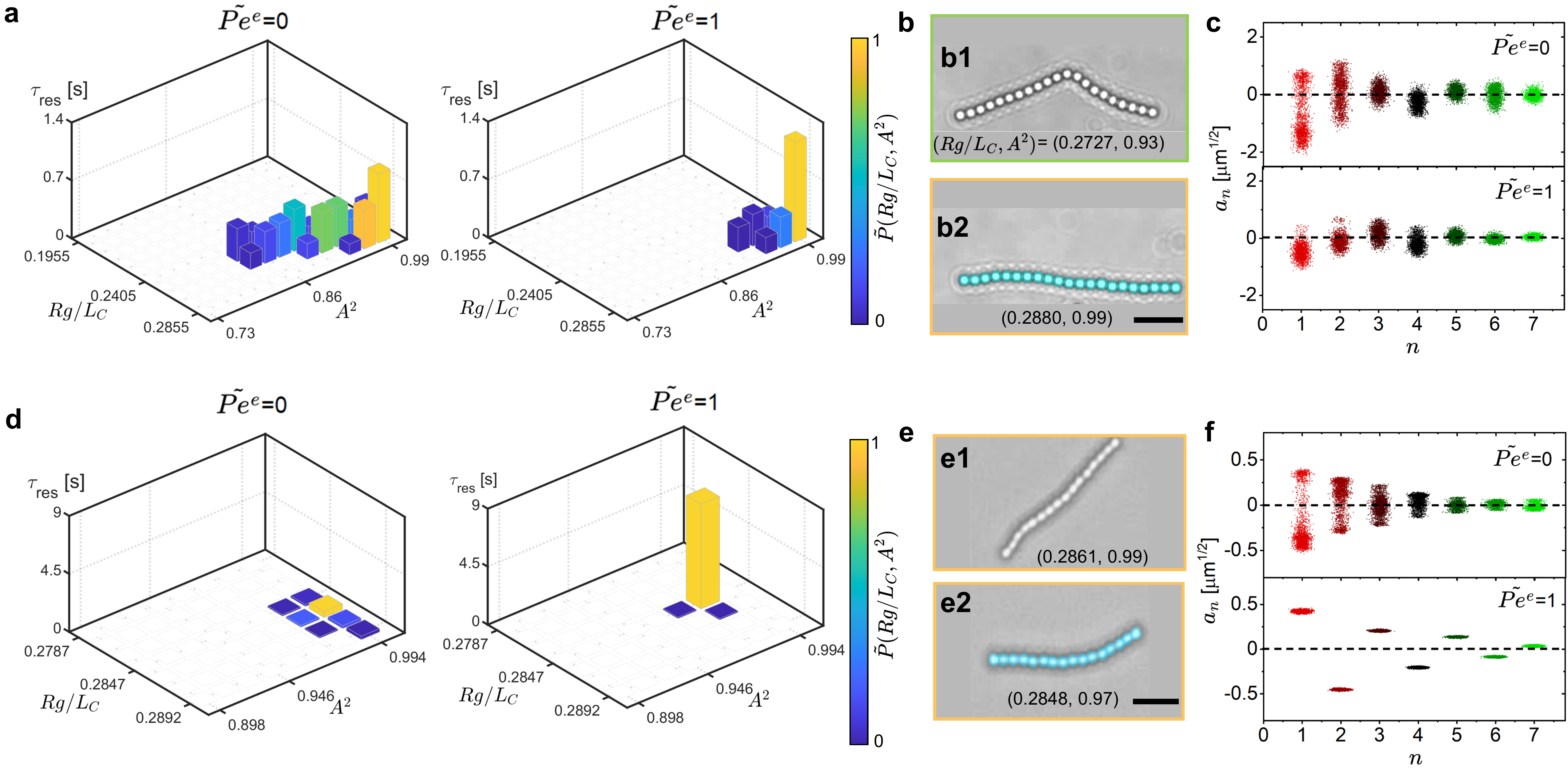}
\caption{\label{fig4} \textbf{Active stiffening of extensile filaments (EFs):} (a) $\mathbb{C}_{n_{1} \times n_{2}}$ of an inherently uniformly cross-linked extensile filament EF1 at Brownian ($\tilde {Pe^e}=0$), and extensile activity ($\tilde {Pe^e}=1$). $\mathbb{C}_{n_{1} \times n_{2}}$ is defined by $Rg/L_C \in [0.11,0.29]$ (x-axis) and $A^2 \in [0.50,1.00]$ (y-axis), divided into $ (n_{1} \times n_{2})$ bins with x-bin width $=0.009$, and y-bin width $=0.02$ . (b) Snapshots of representative conformations of EF1 with corresponding $Rg/L_C$ and $A^2$ values. b1 is a representative conformation at $\tilde {Pe^e}=0$ and b2 the most probable conformations at $\tilde {Pe^e}=1$. (c) $a_n$ for EF1 for all n for both Brownian and  $\tilde {Pe^e}$=1. Filaments with extensile active force show a drastic reduction in the $a_n$ sampled compared to their Brownian counterparts, consistent with conformational maps. EF1 demonstrates uniform stretching into straight conformations under extensile forces. (d) $\mathbb{C}_{n_{1} \times n_{2}}$ of an inherently non-uniformly cross-linked extensile filament EF2 at $\tilde {Pe^e}=0$ and $\tilde {Pe^e}=1$, showing the most probable state to be a non-linear conformation. $\mathbb{C}_{n_{1} \times n_{2}}$ is defined by $Rg/L_C \in [0.26,0.29]$ (x-axis) and $A^2 \in [0.70,1.00]$ (y-axis), divided into $ (n_{1} \times n_{2})$ bins with x-bin width $=0.0015$, and y-bin width $=0.012$ (e) Representative snapshots of the EF2 conformations with corresponding $Rg/L_C$ and $A^2$ for $\tilde {Pe^e}=0$ in e1 and for $\tilde {Pe^e}=$ in e2. The microscopic images confirm that the most probable conformation is a non-linear one. (f) $a_n$ for all n for EF2. The distribution of $a_n$ shows that EF2 attains a non-linear (curved) yet stiff steady-state conformation under extensile activity. In (a) and (d), the colorbar represents $\tilde {P(Rg/L_C, A^2)}$. The color used for the boundaries of images in (b) and (e) corresponds to the $\tilde {P(Rg/L_C, A^2)}$ from the colorbar in (a) and (d), respectively. The scale bar for images in (b) and (e) corresponds to $5 \mu m$. }
\end{figure*}
Finally, we investigate how the interplay between activity and elasticity influences the steady-state conformational properties of CFs by studying filaments with varying intrinsic bending modulus $B$ (Supplementary Figure S5, Supplementary Table 2). Independent of $B$, we find that the contractile activity enables CFs to adopt compact conformations that remain inaccessible to their passive (Brownian) counterparts (Supplementary Figures S8 and S10). However, since $B$ determines a filament's resistance to bending \cite{powers2010dynamics}, the range of accessible conformations narrows as $B$ increases at a fixed $\tilde{Pe^c}$ indicating suppression of conformations with increasing stiffness (Supplementary Figures S8 and S10). To quantitatively capture this competition between activity and elasticity, we consider the scaling of the mean squared local curvature, $\langle \overline{K^2} \rangle_{t}$, measured over a length scale  $\Delta s$, with the contractile active force as  $\sim \eta u_c \Delta s$. This leads to a relation 
\begin{equation}
\langle \overline{K^2} \rangle_{t} \sim \left(\frac{2\eta u_c\Delta s  }{B}\right)^{1}
\label{e1}
\end{equation}
which suggests that the compactness of the filament scales linearly with the ratio of activity to elasticity (supplementary Note 9). Remarkably, for all tested CFs spanning a range of bending moduli ($B$) and activity levels ($\tilde {Pe^c}$), the data for $\langle \overline{K^2} \rangle_{t}$ collapse onto a single linear curve validating equation (\ref{e1}) (Fig. \ref{fig3}d). This collapse unequivocally demonstrates that the competition between contractile active forces and filament elasticity governs the degree of compactness in the steady-state conformations of active CFs.\\\\
\textit{\textbf{Filaments with extensile monomers, EFs}}: 
Next, we study the conformational changes of active filaments composed of extensile monomers (EFs). Unlike CFs, extensile monomers generate source-like EHD flows, which induce effective repulsive interactions between monomers along the filament. In this case, both activity-induced repulsion and filament elasticity cooperate to drive the filament toward an extended, linear conformation. As a result, we find that EFs exhibit activity-induced stiffening, evolving toward a fixed, steady-state shape. $\mathbb{C}_{n_{1} \times n_{2}}$ of EFs reveals a reduced number of distinct conformations compared to their Brownian counterparts (Fig.\ref{fig4}a, and d). Additionally, we find that $\tau_{res} \approx 8s$ of the most-probable conformation (EF2) increases markedly under active conditions and the distribution of 
$a_{n}$s becomes narrower (Fig.\ref{fig4}c and f), both consistent with the emergence of a stable, deterministic steady state. Remarkably, we find that the fixed conformation adopted by an EF in the steady state can be tailored by our filament synthesis. While EF1 adopts a linear steady-state conformation under activity (Fig.\ref{fig4}a, b, and c Supplementary Movies 2 and 8), EF2 that has spatially heterogeneous cross-linking settles into a reproducible non-linear conformation
(Fig.\ref{fig4}d, e, and f Supplementary Movies 3 and 9). Thus, Figure \ref{fig4} establishes that under active forces, EFs stiffen adopting a fixed conformation and the differences in their internal filament structure (heterogenous cross linking density over the filament length), govern the shape of the fixed conformation (Supplementary Figure S9 and S10, Supplementary Movies 2, 3, 8, 9, and 10). 
Our results demonstrate that EHD extensile flows induce local stiffening that can propagate into global filament stretching or a stabilized non-linear shape (Supplementary Movies 2, 3, 8 - 10). Strinkingly similar behaviour has been observed in recent studies of active filaments, including those modelling biological polymers such as chromatin, where monomers generating extensile dipolar flows lead to filament stretching and/or the emergence of stable non-linear conformations  \cite{saintillan2018extensile,mahajan2022self,kumar2024emergent} further emphasizing the relevance of our findings.
\begin{figure*}[ht]

\centering
\includegraphics[width=1\linewidth]{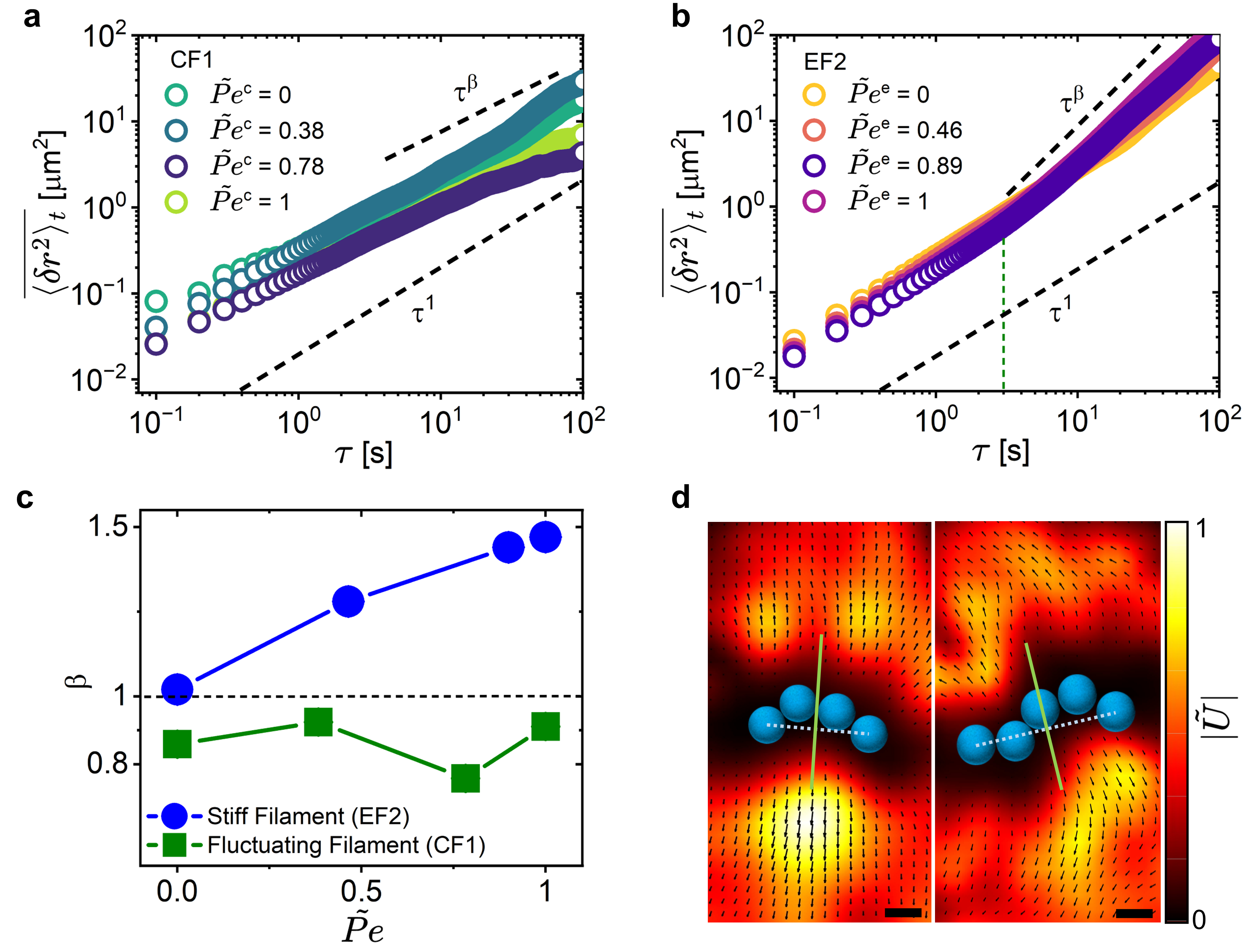}
\caption{\label{fig5} \textbf{Transport of active colloidal filaments:}
(a), and (b) show the monomer averaged mean squared displacement (MSD) $\overline{\langle \delta r^2 \rangle _ {t}} $ for CF1 at different $\tilde{Pe^c}$, and for EF2 at different $\tilde{Pe^e}$, respectively. 
(c) $\beta$, the long time exponent ($\sim \tau >3s$) of MSD, $\overline{\langle \delta r^2 \rangle _ {t}} \propto \tau ^ {\beta}$ as a function of $\tilde {Pe}$, for CF1 (green squares), and EF2 (blue circles). The time for extracting $\beta$ is chosen based on EF2 which shows a transition from diffusive to superdiffusive behavior beyond this time scale (green dotted vertical line in (b)). (d) The mean EHD flow field generated by immobilized EFs via $\mu$PIV for two distinct non-linear conformations showing emergent asymmetric flow field. The flow field is obtained by averaging 100 frames with a fixed conformation for $t \approx 3.12s$ in a steady state. The dotted white line is the end-to-end vector of the conformation, and the green solid line is orthogonal to the end-to-end, shown as guides to the eye. The color bar is the normalized velocity of the flow magnitude, $|\tilde U| = |U|/|U|_{max}$ (here $|U|_{max}=4.75 \mu m /s$). Scale bars = $2 \mu m$.}
\end{figure*}
\section{Transport of active semi-flexible filaments}
To elucidate the role of mean conformation and conformational fluctuations in determining the transport behavior of active filaments we characterize filament motion through the mean-squared displacement (MSD), $\overline{\langle\delta r_{i}^2\rangle_{t}} \propto \tau ^{\beta}$. We find that CFs which exhibit fluctuating steady states display diffusive behavior across all values of $\tilde {Pe^c}$ (Fig\ref{fig5}a, Supplementary Fig. S13a). In contrast, EFs that adopt fixed nonlinear steady-state conformations, exhibit super-diffusive behavior (Fig\ref{fig5}b, Supplementary Fig.S13b). Direct comparison of $\beta$, the long time exponent, confirms this distinction: for EFs $\beta > 1$ for all non-zero $\tilde {Pe^e}$, while for CFs $\beta \approx 1$ even under the largest active force applied (Fig.\ref{fig5}c). Further analysis reveals that for EFs, filaments with odd symmetry conformations show predominantly in-plane rotation, whereas those with even symmetry conformation exhibit translational motion (Supplementary Movie 11), and a linear filament shows no propulsion (Supplementary Figure S13c). To probe the underlying mechanism, we conducted $\mu$PIV measurements on immobilized EFs with odd and even symmetries that show that their self-propulsion arises from emergent asymmetric EHD flow fields around the filament. Odd and even symmetries of the observed flow fields result in rotational and translational motion, respectively (Fig. \ref{fig5}d, Supplementary Movie 12), in agreement with theoretical predictions \cite{jayaraman2012autonomous,ganguly2023going}. Interestingly, while similar asymmetric EHD flows are observed around immobilized nonlinear CFs (Supplementary Figure S12, Supplementary Movie 13), these filaments, when freely suspended do not exhibit enhanced propulsion but instead undergo large conformational fluctuations (Supplementary Figure S8, S10, and S13a). To decouple the effects of shape and shape fluctuations on transport, we synthesized rigid, nonlinear CFs and find that they self-propel, revealing that conformational fluctuations indeed suppress propulsion of semi-flexible CFs (Supplementary Movie 14). Together, these findings reveal that while nonlinear conformations are necessary to generate asymmetric flows in both CFs and EFs, only filaments with persistent shapes such as stiff EFs with long conformational residence times ($\tau_{\text{res}} \sim 8$ s) can sustain the asymmetric flows required for super-diffusion. Highly fluctuating CFs ($\tau_{\text{res}} \sim 0.35$ s) fail to maintain such persistent flows and hence propulsion is not observed for fluctuating CFs. Our results highlight that, beyond the nature of the active forces (contractile vs. extensile) and instantaneous conformation, the timescale of conformational dynamics plays a critical role in dictating the transport properties of active filaments.  
\section{Conclusions}
Our work pioneers the experimental realization of synthetic semi-flexible active filaments with tunable emergent behavior driven by monomer-level hydrodynamic forces. Our results show that filament conformational dynamics stem from the interplay between activity and elasticity, leading to fluctuating steady states in contractile filaments (CFs) (Fig.\ref{fig3}) and active stiffening in extensile ones (EFs) (Fig.\ref{fig4}). Our studies reveal that both the mean conformation and the timescale of conformational fluctuations play pivotal roles in determining transport behavior with stiff, nonlinear filaments exhibiting self-propulsion (EF2, EF3, Stiff CF; see Supplementary Movie 14), and continuously fluctuating ones remain diffusive (CF1–CF4). By independently controlling active force generation and intrinsic filament stiffness, we offer a framework for designing and understanding active filaments based on activity–rigidity balance. Furthermore, our modular system lays the groundwork for future investigation of fundamental studies of active filaments including their collective behavior and provides a versatile platform for development of adaptive, flexible microbots.

\newpage


\section{Competing interests}
The authors declare no competing interests.\\

\section{Acknowlegements}
The authors thank Narayanan Menon for helpful discussions. H.N.K. thanks the University of Massachusetts Amherst for the start-up funding.\\

\section{Author contributions}
B.B. \& P.M. performed the experiments

\bibliography{Ref}

\end{document}